# An Actively Accreting Massive Black Hole in the Dwarf Starburst Galaxy Henize 2-10

Amy E. Reines[1], Gregory R. Sivakoff[1], Kelsey E. Johnson[1,2] and Crystal L. Brogan[2]

[1]*Department of Astronomy, University of Virginia, 530 McCormick Road, Charlottesville, VA 22904, USA*

[2] *National Radio Astronomy Observatory, 520 Edgemont Road, Charlottesville, VA 22904, USA*

**Supermassive black holes are now thought to lie at the heart of every giant galaxy with a spheroidal component including our own Milky Way[1,2]. However, the birth and growth of the first "seed" black holes in the earlier Universe is observationally unconstrained[3] and we are only beginning to piece together a scenario for their subsequent evolution[4]. Here we report that the nearby dwarf starburst galaxy Henize 2-10[5,6] contains a compact radio source at the dynamical centre of the galaxy that is spatially coincident with a hard X-ray source. From these observations, we conclude that Henize 2-10 harbours an actively accreting central black hole with a mass of approximately one million solar masses. This nearby dwarf galaxy, simultaneously hosting a massive black hole and an extreme burst of star formation, is analogous in many ways to galaxies in the infant Universe during the early stages of black hole growth and galaxy mass assembly. Our results confirm that nearby star-forming dwarf galaxies can indeed form massive black holes, and by implication so can their primordial counterparts. Moreover, the lack of a substantial spheroidal component in Henize 2-10 indicates that supermassive black hole growth may precede the build-up of galaxy spheroids.**

The starburst in Henize 2-10, a relatively nearby (9 Mpc, ~30 million light years) blue compact dwarf galaxy, has attracted the attention of astronomers for decades[6-10]. Stars are forming in Henize 2-10 at a prodigious rate[8,11,12] that is ten times that of the Large Magellanic Cloud[13] (a satellite galaxy of the Milky Way), despite the fact that these two dwarf galaxies have similar stellar masses[14,15,16] and neutral hydrogen gas masses[7,17]. Most of the star formation in Henize 2-10 is concentrated in a large population of very massive and dense "super star clusters", the youngest having ages of a few million years and masses of one hundred thousand times the mass of the Sun[6]. The main optical body of the galaxy has an extent less than 1 kpc (~3000 light-years) in size and has a compact irregular morphology typical of blue compact dwarfs (Fig. 1).

We observed Henize 2-10 at centimetre radio wavelengths with the Very Large Array (VLA) and in the near-infrared with the *Hubble Space Telescope (HST)* as part of a large-scale panchromatic study of nearby dwarf starburst galaxies harbouring infant super star clusters[18,19,20]. A comparison between the VLA and *HST* observations drew our attention to a compact (< 24 pc × 9 pc) central radio source located between two bright regions of ionized gas (Fig. 2). These data exclude any association of this central radio source with a visible stellar cluster (Fig. 3; see Supplementary Information for a discussion of the astrometry). Furthermore, the radio emission from this source has a significant non-thermal component ($\alpha \sim -0.4$, $S_\nu \propto \nu^\alpha$) between 4.9 GHz and 8.5 GHz, as noted in previous studies of the galaxy[9]. An archival observation of Henize 2-10 taken with the *Chandra X-ray Observatory* reveals that a point source with hard X-ray emission is also coincident (to within the position uncertainty) with the central non-thermal radio source[10]. Typically, even powerful non-nuclear radio and X-ray sources (e.g. supernova remnants and active X-ray binaries) are at least an order of magnitude fainter than the central source in Henize 2-10 (see Supplementary Information). In contrast, the radio and hard X-ray luminosities of the central source in Henize 2-10, as well as their ratio, are similar to known low-luminosity active galactic nuclei powered by accretion onto a massive black hole[21].

The central, compact, non-thermal radio source in Henize 2-10 is also coincident with a local peak in Paα and Hα emission and appears to be connected to a thin quasi-linear ionized structure between two bright and extended regions of ionized gas. This morphology is tantalizingly suggestive of outflow (Fig. 2). While we cannot conclusively determine whether or not this linear structure is physically connected to the brightest emitting regions with the data in hand, ground-based spectroscopic observations[22] confirm a coherent velocity gradient along the entire ionized gas structure seen in Figure 2 consistent with outflow or rotation. Moreover, a comparison between the central velocity of this ionized gas structure and the systemic velocity of the galaxy derived from observations of neutral hydrogen gas that rotates as a solid body[7] indicates that the position of the non-thermal radio source is consistent with the dynamical centre of the galaxy.

While compact radio and hard X-ray emission at the centre of a galaxy are generally good indicators of accretion onto a massive black hole[21], we have considered alternative explanations for the data. As discussed at length in the Supplementary Information, it is extremely unlikely that the nuclear source in Henize 2-10 is one or more supernova remnants, more recently created supernovae, stellar mass black hole X-ray binaries, or some combination of these phenomena. Briefly, X-ray binaries are too weak in the radio, supernova remnants are too weak in hard X-rays, and young compact radio supernovae are ruled out by observations using Very Long Baseline Interferometry[23]. While it may be possible to account for the radio and X-ray luminosities of the nuclear source with just the right combination of the aforementioned phenomena, the probability of such a coincidence is exceedingly low (see Supplementary Information). On the contrary, the radio and hard X-ray luminosities of the central source in Henize 2-10 are well within the range of known low-luminosity active galactic nuclei[21].

In addition to ruling out young compact supernovae, the non-detection of the nuclear radio source at higher resolution (~0.5 × 0.1 parsec) using Very Long Baseline Interferometry[23] may also seemingly rule out the presence of an actively accreting massive black hole. However, Seyfert nuclei with steep radio spectra ($\alpha \sim < -0.5$, $S_\nu \propto \nu^\alpha$) commonly exhibit this "missing flux" phenomenon when observed at increasingly higher spatial resolution[24]. In these active galactic nuclei, as much as ~90% of the radio emission is absent on parsec scales and is expected to be dominated by extended low-surface-brightness features on larger scales, such as jets. This is in contrast to Seyferts with flat or positive radio spectra ($\alpha \sim > 0$) and elliptical radio galaxies in which the radio emission is concentrated in a compact core. The nuclear radio source in Henize 2-10 has a spectral index ($\alpha \sim -0.4$) similar to Seyferts that are known to have reduced flux densities on parsec scales. Therefore, we do not consider the non-detection of the nuclear radio source at very high resolution to be incompatible with the presence of an active galactic nucleus in Henize 2-10.

We conclude that an actively accreting massive black hole is the most feasible explanation for the nuclear source in Henize 2-10. The compact radio and hard X-ray luminosities are consistent with the observed correlation for active galactic nuclei and we therefore estimate the mass of the black hole in Henize 2-10 using the so-called "fundamental plane of black hole activity"[25]. This empirical relationship relating black hole mass to the emitted compact radio and hard X-ray luminosities, spanning nine orders of magnitude in black hole mass, is given by the equation $\log L_R = 0.60 \log L_X + 0.78 \log M + 7.33$, where $L_R$ is the radio luminosity at 5 GHz in erg s$^{-1}$, $L_X$ is the 2-10 keV X-ray luminosity in erg s$^{-1}$, and $M$ is the mass of the black hole in solar masses. Using the observed radio luminosity of $7.4 \times 10^{35}$ erg s$^{-1}$ at 4.9 GHz and the X-ray luminosity of $2.7 \times 10^{39}$ erg s$^{-1}$ in the 2-10 keV band, we calculate $\log(M/M_\odot) = 6.3 \pm 1.1$ for the black hole in Henize 2-10. The region in which the gravitational influence of a one million solar mass black hole dominates that of the host galaxy subtends a very small angle on the sky at the distance of Henize 2-10 (< 1 arcsecond for velocity dispersions > 10 km s$^{-1}$). Thus, it is not surprising that kinematic studies of Henize 2-10 have not previously revealed the presence of the black hole at its centre.

Few dwarf galaxies are currently known to host massive black holes[26,27], however the discovery of an active nucleus in Henize 2-10 opens up a new realm in which to search for local analogues of primordial black hole growth (i.e. dwarf starburst galaxies). While recent searches[28,29] have revealed growing numbers of nuclear black holes with masses similar to that in Henize 2-10, the host galaxies of these objects have very different properties than Henize 2-10. Most notably, they are not actively forming stars and have regular morphologies of disks and spheroids with well-defined optical nuclei[29,30]. Moreover, the majority of the black holes detected in these systems are radiating at high fractions of their Eddington limits[27,28,29], suggesting the black holes are presently undergoing rapid growth. In contrast, the low-luminosity active galactic nucleus in Henize 2-10 is currently radiating significantly below its Eddington limit (~$10^{-4}$ assuming an X-ray bolometric correction of 10, see Supplementary Information), suggesting a different evolutionary state.

The results presented here have broad implications for our understanding of the evolution of galaxies and their central black holes. The concurrent black hole growth and extreme starburst in Henize 2-10 likely resembles the conditions in low-mass high-redshift galaxies during the early phases of galaxy assembly when interactions and mergers were common. Indeed, Henize 2-10 shows signs of having undergone an interaction, including tidal tail-like features in both its gaseous[7] and stellar distributions (Fig. 1). Additionally, it is intriguing that the massive black hole in Henize 2-10 does not appear to be associated with a bulge, a nuclear star cluster, or any other well-defined nucleus. This unusual property may reflect an early phase of black hole growth and galaxy evolution that has not been previously observed. If so, this implies that primordial seed black holes could have pre-dated their eventual dwellings, thereby constraining theories for the formation mechanisms of massive black holes and galaxies.

**Supplementary Information** is linked to the online version of the paper at www.nature.com/nature.

**Acknowledgements** A.R. is grateful for many discussions on this work, in particular with Mark Whittle, Jim Ulvestad, Miller Goss, Sheila Kannappan, Jenny Greene, Ryan Hickox, Aaron Evens, Bob O'Connell, Roger Chevalier, Anil Seth, Elena Gallo, Scott Ransom, Leslie Hunt and Jake Simon. A.R. acknowledges support from a NASA Earth and Space Science Fellowship, and the University of Virginia through a Governor's Fellowship and a Dissertation Acceleration Fellowship. G.S. acknowledges support for this work by NASA through the *Chandra X-ray Observatory* Center, which is operated by the Smithsonian Astrophysical Observatory for and on behalf of NASA. K.J. acknowledges support from the NSF through a CAREER award and the David and Lucile Packard Foundation through a Packard Fellowship. Support was provided by NASA through a grant from the Space Telescope Science Institute, which is operated by the Association of Universities for Research in Astronomy, Inc. The National Radio Astronomy Observatory is a facility of the National Science Foundation operated under cooperative agreement by Associated Universities, Inc. This research has made use of data obtained from the Hubble Space Telescope and Chandra X-ray Observatory Data Archives.

**Author Contributions** A.R. reduced the *HST*/NICMOS images, synthesized the multi-wavelength data, and led the analysis, interpretation, and writing of the paper. G.S. analyzed the *Chandra* data, and helped with the interpretation and writing of the paper. K.J. led the *HST*/NICMOS and VLA proposals. K.J. and C.B. reduced and analyzed the VLA data. All authors discussed the results and presentation of the paper.

**Author Information** Reprints and permissions information is available at www.nature.com/reprints. The authors declare that they have no competing interests as defined by Nature Publishing Group, or other interests that might be perceived to influence the results and discussion reported in this article. Correspondence and requests for materials should be addressed to A.R. (areines@virginia.edu).

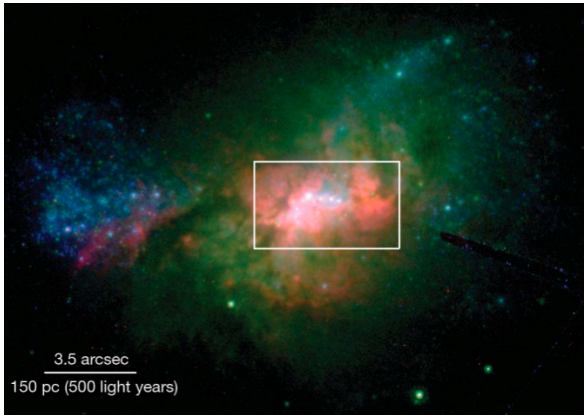

**Figure 1: Henize 2-10.** Henize 2-10 is a blue compact dwarf galaxy hosting a concentrated region of extreme star formation. Using H$\alpha$[8] and 24 micron[11] fluxes from the literature, we estimate a star formation rate[12] of 1.9 $M_\odot$ yr$^{-1}$ assuming all of the emission is from the starburst and that the contribution from the active nucleus is negligible. We estimate that Henize 2-10 has a stellar mass of $3.7 \times 10^9$ $M_\odot$ from the integrated 2MASS $K_s$-band flux[14,15]. Neutral hydrogen observations of Henize 2-10 indicate a solid body rotation curve typical of dwarf galaxies with a maximum projected rotational velocity of 39 km s$^{-1}$ relative to the systemic velocity of the galaxy[7]. These observations also indicate a dynamical mass of ~$10^{10}$ solar masses within 2.1 kpc[7]. The main optical body of the galaxy, shown here, is less than 1 kpc across. Henize 2-10 shows signs of having undergone an interaction, including tidal-tail like features in both its gaseous[7] and stellar distributions (seen here). In this three-color *HST* image of the galaxy, we show ionized gas emission in red (H$\alpha$) and stellar continuum in green (~ *I*-band, 0.8 micron) and blue (~ *U*-band, 0.3 micron). These archival data were taken with Wide Field and Planetary Camera 2 (H$\alpha$) and the Advanced Camera for Surveys (*U*- and *I*-band). The white box indicates the region shown in Figures 2 and 3.

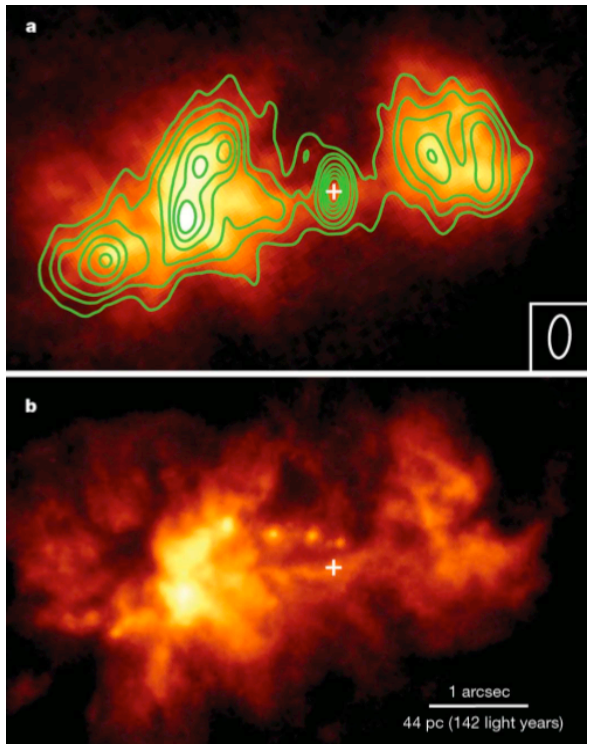

**Figure 2: The Active Nucleus and Ionized Gas in Henize 2-10.** The overall morphology of the radio emission (green contours) in the central region of Henize 2-10 matches that of the ionized gas detected with *HST* (colour images), suggesting a shared origin. The active nucleus (plus symbol) is detected as a non-thermal nuclear VLA radio source coincident with a *Chandra* point source with hard X-ray emission. The nuclear source is also coincident with a local peak of ionized gas emission and appears connected to the thin quasi-linear structure between the two bright and extended regions of ionized gas, suggestive although not proof of outflow. The central source has 4.9 GHz and 8.5 GHz radio luminosities of $7.4 \times 10^{35}$ erg s$^{-1}$ and $1.0 \times 10^{36}$ erg s$^{-1}$, respectively. The X-ray luminosity of the central source in the 2-10 keV band is $\sim 2.7 \times 10^{39}$ erg s$^{-1}$. The accretion rate of the $2 \times 10^{6}$ M$_\odot$ black hole is $\sim 5 \times 10^{-6}$ M$_\odot$ yr$^{-1}$ assuming an X-ray bolometric correction of 10 and an accretion efficiency of 0.1. (a) Narrowband imaging with *HST*/NICMOS was used to trace the ionized gas in Henize 2-10 using the Paα hydrogen recombination line at 1.87 microns. Continuum emission was removed using a neighbouring off-line narrowband filter. VLA 8.5 GHz (3.5 cm) radio contours are over-plotted in green and the active galactic nucleus is marked with a plus symbol. Contour levels are 9, 13, 17, 25, 33, 41, 49 × the r.m.s. noise (12 µJy beam$^{-1}$). The beam is shown in the lower right corner. (b) Optical narrowband imaging of the H$\alpha$ hydrogen recombination line at 0.66 microns yields a higher-resolution view of the ionized gas in Henize 2-10. The continuum has not been subtracted in this archival image, leaving young star clusters also visible.

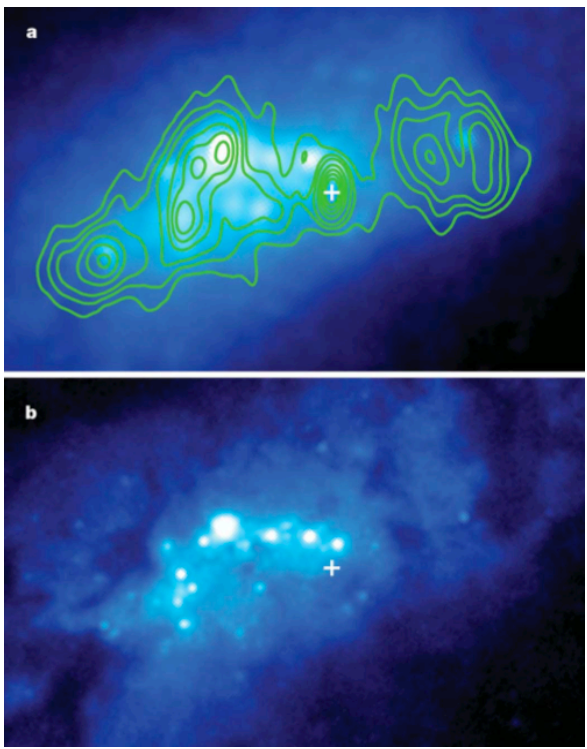

**Figure 3: Young super star clusters in Henize 2-10.** The overall structure of the radio emission (green contours) differs markedly from the distribution of star clusters in the centre of the galaxy (colour images). In particular, the non-thermal nuclear radio source does not have a detectable counterpart in these broadband continuum images (plus symbol) excluding any association with a visible star cluster. (a) A near-infrared (~ *K*-band) image of the central region of Henize 2-10 over-plotted with the same radio contours from Figure 2. *HST*/NICMOS was used to observe the galaxy through a broadband filter centred at 2.1 microns, which primarily traces the distribution of stellar light. (b) A higher-resolution view of the star clusters is shown in this archival 0.8 micron (~ *I*-band) broadband image. The field of view is the same as in Figure 2.

## Observations and Data Reduction

A summary of the observations used in this work is presented in Supplementary Table 1. A detailed description of the observations is given below.

**Near-IR and Optical Imaging with *HST***

High-resolution near-infrared *Hubble Space Telescope* (*HST*) observations of Henize 2-10 were obtained with the Near Infrared Camera and Multiobject Spectrometer (NICMOS) using the NIC 2 Camera. Images through the F187N, F190N, and F205W filters were taken on 2006 October 1. The F187N and F190N narrowband filters cover the Pa$\alpha$ hydrogen recombination line (rest wavelength of 1.87$\mu$m) and the neighboring continuum, respectively. At the redshift of Henize 2-10 ($z = 0.002912$), the Pa$\alpha$ line falls near the centre and maximum throughput of the F187N filter. The broadband F205W filter ($\sim K$-band) is centred at approximately 2.1$\mu$m. Four dithered sub-exposures were taken in each filter with total on-source integration times of 896 s (F187N), 960 s (F190N) and 480 s (F205W). In addition, off-source observations of equal time were obtained in order to mitigate the effects of the thermal background.

The data were processed following the procedure outlined in Reines et al. (2008b)[19]. The final images have a plate scale of $\sim 0\farcs 0375$ pixel$^{-1}$ and are essentially diffraction limited with spatial resolutions of $\sim 0\farcs 2$. The F187N and F190N images were used to construct a continuum-subtracted Pa$\alpha$ flux image. The image units of DN s$^{-1}$ were converted to flux densities using the inverse sensitivity PHOTFLAM header keywords. The continuum image (F190N) was then subtracted from the line image (F187N), which in turn was multiplied by the width of the F187N filter.

In addition to the new NICMOS observations presented here, we also utilize archival optical observations of Henize 2-10. Images through the F814W ($\sim I$-band, 0.8 $\mu$m) and F330W ($\sim U$-band, 0.3 $\mu$m) filters were obtained with the Advanced Camera for Surveys (ACS) High Resolution Channel (HRC) on 2006 March 12 under Proposal ID 10609 (W. Vacca). We retrieved the pipeline-produced, calibrated and drizzled images from the *HST* archive. The HRC has a plate scale of $\sim 0\farcs 027$ pixel$^{-1}$ and the resolution of the images of Henize 2-10 are in the range $\sim 0\farcs 06$ - $0\farcs 08$. We also obtained archival narrowband images of Henize 2-10 taken through an H$\alpha$ filter. The observations were taken on 1997 April 3 with the Wide Field and Planetary Camera 2 (WFPC2) under Proposal ID 6580 (P. Conti). The F658N filter was used and the galaxy was placed on the PC chip with a plate scale of $\sim 0\farcs 05$ pixel$^{-1}$. We combined four pipeline-produced calibrated sub-exposures and rejected cosmic rays using custom IDL programs. The final F658N image has a resolution of $\sim 0\farcs 13$. Since no narrowband continuum image was available for subtraction, the F658N image contains both line and continuum emission.

**Radio Continuum Observations with the VLA**

We obtained new X-band (8.5 GHz, 3.5 cm) and C-band (4.9 GHz, 6.2 cm) observations of

Henize 2-10 in December of 2004 with the Very Large Array (VLA)*. These observations were taken in the A-array and utilized the the link to the Pie Town Antenna in the Very Long Baseline Array, resulting in an increased spatial resolution roughly a factor of two better than possible with the VLA alone. The resulting data are the highest resolution and most sensitive VLA observations of Henize 2-10 available to date.

The data were calibrated using the Astronomical Image Processing System (AIPS) software package. The flux density scale was calibrated using the standard sources 0713+438 and 3C286 for both the 3.5 cm and 6.2 cm data. The target and phase calibrator (0836-202) were observed using "fast switching mode" at both frequencies: for the 3.5 cm data, the phase calibrator was observed for 2 minutes before and after every 10 minutes on Henize 2-10, and for the 6.2 cm data the phase calibrator was observed for 2 minutes before and after every 14 minutes observing Henize 2-10. A total of $\sim 4$ hours and $\sim 5$ hours of integration time were spent on Henize 2-10 at 3.5 cm and 6.2 cm, respectively.

Given the nature of interferometric observations, it is not possible to obtain identical $u-v$ coverage at each wavelength. However, in order to obtain images at each wavelength with well-matched spatial sensitivity, we endeavor to attain relatively well matched synthesized beams using three steps: first, the $uv$-range of the observations is restricted to values greater that 6.5 k$\lambda$ at each wavelength, which will result in approximately similar sensitivity to large spatial scales; second, we vary the weighting used in the imaging process via the "robust" parameter (robust = 5 invokes purely natural weighting, while a robust = -5 invokes purely uniform weighting); and third, the resulting images are convolved to identical resolution elements. The resulting 3.5 cm image was created using robust=5 and the 6.2 cm image with robust = -1, which resulted in synthesized beams of $0\farcs44 \times 0\farcs21$ and $0\farcs55 \times 0\farcs15$, respectively. The final images were convolved to $0\farcs55 \times 0\farcs21$, which corresponds to a FWHM of $\sim 24$ pc$\times 9$ pc at the distance of Henize 2-10. At this resolution, the central source in Henize 2-10 is consistent with being point-like.

We also re-reduced and analyzed existing archival radio observations of Henize 2-10 for comparison. In this process, we found that the 6.2 cm data reported in Johnson & Kobulnicky (2003)[9] that was observed in 1995 had flux densities about half those of the new data for all of the compact sources in Henize 2-10. This discrepancy is likely due to the 1995 data having being taken during poor atmospheric conditions, which resulted in significant decorrelation of the signal, and hence low flux densities. Consequently, we have opted to exclude the 1995 data from the analysis presented in this paper.

Flux densities were measured using the VIEWER program in the CASA software package. Identical apertures were used at each wavelength with an area of approximately twice the size of the synthesized beam (resulting in an ellipse of $\sim 1'' \times 0\farcs5$). Given the substantial diffuse emission in the area, background estimates were also made using the source-free areas to the north and south of the central object, and the background values were subtracted from the source aperture flux density to obtain the final values of 1.25 mJy and 1.55 mJy at 3.5 cm and 6.2 cm, respectively. We estimate the uncertainty in the relative flux densities to be $\sim 10 - 20\%$ based on the absolute flux calibration.

---

*The National Radio Astronomy Observatory is a facility of the National Science Foundation operated under cooperative agreement by Associated Universities, Inc.

*Chandra* **X-ray Data**

Henize 2-10 was observed for 19755 s with the Advanced CCD Imaging Spectrometer (ACIS) detector of the *Chandra X-ray Observatory* (CXO) on 2001 March 23 (Observation 2075). Although previous analyses have discussed the diffuse gas and the central source of Henize 2-10 revealed in X-rays[10,31], we undertook an independent analysis of the X-ray point sources to verify the association of the central source with the non-thermal radio source and determine its X-ray properties. Our primary analysis used CIAO 4.1[32] with CALDB 4.1.1 and NASA's HEASOFT 6.6.2[†].

Throughout the observation, Henize 2-10 was located on the ACIS-7 (S3) chip and no background flares were observed. To detect X-ray point sources, we supplied the CIAO WAVDETECT routine with an image of the S3 chip over the 0.3–6 keV range and a mono-energetic exposure map at 1 keV. We used this wavelet detection algorithm to identify sources with $\sqrt{2}$ scales ranging from 1 to 32 pixels ($\sim 0\rlap{.}''5$–$16''$) with a source detection threshold of $10^{-6}$. We detected 35 X-ray point sources, while only expecting $\lesssim 1$ false source due to a statistical fluctuation in the background.

Since CIAO WAVDETECT does not take into account the shape of the X-ray PSF, we refined the source positions using ACIS Extract 2009-01-27[33]. For each source, we also used ACIS Extract to create a source extraction region set to encircle 90% of the flux in the X-ray PSF, as well as a circular masking region and a local background region. We then estimated the statistical error in the position using the net counts for each source[34].

The central X-ray source in Henize 2-10 is associated with the non-thermal radio source (see below), and we hereafter only consider this X-ray source with $\sim 180$ net counts. We extracted spectra for the source and its background in the 0.5–10 keV range (Supplementary Figure 1). The source spectrum was grouped to have 16 counts per bin, and all spectra were fitted using the Cash statistic[35]. As noted previously[10,31], the source emission has peaks at both soft energies ($\lesssim 1$ keV) and hard energies ($\sim 3$ keV). The peak at hard X-ray emission likely points to a non-thermal process that is moderately affected by absorption. Since the source lies near the peak of the diffuse interstellar medium in Henize 2-10[10], some of its soft energy counts may arise from diffuse gas that undergoes weaker absorption, even after the subtraction of a local background. Alternatively, the soft emission could be intrinsic to the source; however, for most spectral models the observed spectral shape implies that there must be different absorption column densities for the soft and hard components. All models were fit to spectra of the source and nearby backgrounds using XSPEC

---

[†]See http://heasarc.gsfc.nasa.gov/docs/software/lheasoft/.

12.5.

For our fiducial source model, we assumed the spectrum was due to a combination of absorbed diffuse gas and an absorbed power-law, tbabs$_{\rm Gal}$ * (tbabs$_{\rm ISM}$ * vmekal$_{\rm ISM}$ + tbabs$_{\rm pow}$ * pow). We adopted updated abundances and photoelectric absorption cross-sections[36,37] for use with the Tuebingen-Boulder ISM absorption model (tbabs). The power-law spectra, $N(E) \propto E^{-\Gamma}$, is what one might expect from a hard-state X-ray binary or an active galactic nucleus; similar spectra have also been used to model supernova remnants like the Crab nebula. To model the contribution from diffuse gas, we adopted a fit to the diffuse gas of the entire galaxy by Kobulnicky & Martin (2010)[10], an optically thin thermal plasma model (0.64 keV) with independent abundances of $\alpha$ (0.77 Solar) and Fe-peak (0.29 Solar) elements (i.e., a vmekal model).

Since the diffuse gas fit from the literature was over a large region of the galaxy and used the wabs absorption model, the absorption column density near the source may differ. We therefore first fit this model to spectra extracted from a region within $\sim 5''$ of the central source, excluding the masking regions of all sources and allowing the model normalization and local absorption along the line-of sight to vary. This led to a larger absorption ($N_{H,\rm ISM} = 1.3 \times 10^{21}$ cm$^{-2}$ on top of $N_{H,\rm Gal^{38}} \equiv 9.7 \times 10^{20}$ cm$^{-2}$) than reported in Kobulnicky & Martin (2010)[10]. We adopted our larger absorption column densities, but allowed the normalization to vary for modeling the excess ISM inside the source extraction region.

The best-fit model found $N_{H,\rm pow} = (6.3^{+5.5}_{-3.6}) \times 10^{22}$ cm$^{-2}$ and $\Gamma_{\rm pow} = 1.66^{+1.16}_{-0.85}$ with a C-statistic of 7.34 for 8 degrees of freedom; the errors reported represent 90% confidence intervals. Since $\sim 16\%$ of realization simulated from the best fit have a lower C-statistic, this suggests the model is an acceptable fit. We note that the normalization of the vmekal model is consistent with expectations for excess diffuse ISM in the source extraction region. To avoid difficulties accounting for the moderately strong absorption at soft energies, as well as placing this source on the fundamental plane of black hole activity, we calculated the observed fluxes and intrinsic luminosities in the 2–10 keV band. Correcting for the encircled energy of the source extraction regions, the observed fluxes from models within the confidence interval are 1.7–2.6 $\times 10^{-13}$ erg cm$^{-2}$ s$^{-1}$, with a best-fit flux of $2.1 \times 10^{-13}$ erg cm$^{-2}$ s$^{-1}$. We note that our lowest flux is consistent with that derived by Kobulnicky & Martin (2010)[10]. After removing Galactic and local absorption, the intrinsic luminosities range from 2.7–3.5 $\times 10^{39}$ erg s$^{-1}$, with a best-fit luminosity of $2.7 \times 10^{39}$ erg s$^{-1}$.

To verify that these numbers are not biased by our choice of fiducial spectra, we tried a variety of other plausible models that might arise from X-ray binaries, active galactic nuclei, supernova remnants, or young supernovae. While the current spectra are not sufficiently accurate to pinpoint the spectral components, they do produce a consistent flux and intrinsic luminosity in the hard 2–10 keV band.

**Astrometry**

To facilitate the comparison of multi-wavelength observations of Henize 2-10, we registered the *HST*, VLA, and CXO data to a common astrometric reference frame defined by the 2MASS catalog[14], which has typical source positions accurate to $\lesssim 0\farcs 1$. More accurate astrometric registration can be achieved by using the mean registration from multiple sources. Before such registration, the absolute positional accuracy of source coordinates are $\sim 1''$, $\sim 0.1''$, and $\sim 0.5''$, for the *HST*, VLA, and CXO observations, respectively.

All *HST* images were co-aligned to an archival WFPC2 F814W image of Henize 2-10 placed on the WF3 chip (P.I. Calzetti) using the bright stellar clusters common to each image. Although the WFPC2/WF3 F814W image has a lower resolution than the ACS/HRC F814W image shown in Figures 2 and 3, its larger field of view is needed to match four point sources located outside of Henize 2-10 with 2MASS sources. This produced an absolute registration accuracy of $\sim 0\farcs 05$ in RA and $\sim 0\farcs 04$ in Dec. Since the radio continuum VLA images have no overlap with 2MASS sources, but have features in common with the *HST* Paα image, we used this newly registered image to determine that the VLA images required an astrometric shift of $\sim 0\farcs 08$ west and $\sim 0\farcs 10$ north, which is reasonable given the absolute astrometric accuracy of the VLA alone. We note that the central non-thermal radio source was not used in the astrometric registration procedure and that the Paα and Hα counterparts to the radio source were discovered after the image registration. Applying these shifts to the original VLA coordinates of the central non-thermal source, we obtain a position of $08^h36^m15\fs117$ in RA and $-26°24'34\farcs07$ in Dec.

To determine the absolute astrometry of our X-ray sources, we similarly matched the X-ray source positions against 2MASS positions, excluding the two X-ray point sources within $5''$ of the non-thermal radio source. Using five sources matched to within their $3\sigma$ positional uncertainties, we determined the absolute astrometric shift, a $\sim 0\farcs 05$ shift west and $\sim 0\farcs 02$ shift south of the original CXO positions. The combined statistical and systematic error of the shift is $0\farcs 12$ in RA and $0\farcs 33$ in Dec. Since the hard central X-ray is close to the optical axis of the CXO observation and has $\sim 180$ net counts, its positional error of $0\farcs 13$ in RA and $0\farcs 33$ in DEC is dominated by the error in absolute astrometry. The hard X-ray source, located at $08^h36^m15\fs114$ in RA and $-26°24'33\farcs75$ in Dec, is coincident with the non-thermal radio source within the $1\sigma$ error on the CXO position.

One potential problem with registering observations taken at different wavelengths at different times is the proper motions of sources that define the reference frame (2MASS sources in this case). However, we note that this would not affect the relative astrometry between the *HST* and VLA images. Therefore, our conclusion that the central non-thermal radio source does not coincide with an optical star cluster remains secure. The same is true for our finding that the radio source is spatially coincident with a local peak in Hα and Paα emission. The main concern is the position of the central non-thermal radio source with respect to the hard X-ray source. The *Chandra* observation was taken in early 2001 and the point sources in that image that were matched to 2MASS sources ($\sim$ epoch 2000 coordinates) all have proper motions less than $\sim 25$ mas yr$^{-1}$. Therefore, the sources defining the reference frame had not moved significantly roughly a year later when the *Chandra* observation was taken. The *HST* image (and therefore the VLA positions) that we tied to 2MASS was taken in early 2008. However, the four stars used in the registration process have proper motions less than ∼5 mas yr$^{-1}$ (or have no measurement). So again, it does not appear that the sources defining the reference frame had moved significantly at the time the *HST* observation was taken.

## Discussion

**Possible Alternative Origins for the Nuclear Source in Henize 2-10**

While we have already shown (in the main paper) that a low-luminosity AGN is a natural explanation for the nuclear source in Henize 2-10, we have also considered whether the nuclear source could have an alternative origin such as one or more X-ray binaries, supernova remnants, more recently created supernovae, or some combination of these objects. A tantalizing interpretation for the morphology of the quasi-linear structure seen in ionized gas emission is outflow from the central source, for which the aforementioned phenomenon would have difficulty accounting for. However, in the discussion that follows, we simply consider the radio and X-ray luminosities of the nuclear source and disregard any possible connection to the elongated ionized gas structure. We also make use of the *ratio* of the radio to X-ray luminosities, $R_X = \nu L_\nu(5 \text{ GHz})/L_X(2-10 \text{ keV})$[39]. For reference, the nuclear source in Henize 2-10 has log $R_X \sim -3.6$. Typical low-luminosity AGN have log $R_X$ between $-2.8$ and $-3.8$[21].

i. Stellar-mass black hole X-ray binaries?

While the most luminous black hole X-ray binaries may be capable of producing the observed hard X-ray emission from the nuclear source in Henize 2-10, these objects have a deficit of radio emission. Using a sample of seven Galactic black hole X-ray binaries[25], we find log $R_X < -5.3$. X-ray binaries are simply too weak in the radio to account for both the X-ray *and* radio properties of the nuclear source in Henize 2-10.

ii. Supernova Remnants?

Multiple supernova remnants could in principle account for the non-thermal radio emission coming from the nuclear source in Henize 2-10; however, these objects have relatively weak X-ray luminosities in the 2-10 keV band. Take Cas A for example, one of the most luminous Galactic supernova remnants in the radio and hard X-rays with an age of ∼300 years. At epoch 2000, Cas A had a 5 GHz luminosity[40,41] of $4.7 \times 10^{34}$ erg s$^{-1}$ and a 2-10 keV luminosity[42] of $2.6 \times 10^{36}$ erg s$^{-1}$, giving log $R_X \sim -1.7$.

We also consider the ultraluminous supernova remnant J1228+441[43] in the Magellanic-type irregular starburst galaxy NGC 4449. At a distance of 3.8 Mpc[44], the 5 GHz luminosity[45] of this supernova remnant in 2002 was $\sim 3.4 \times 10^{35}$ erg s$^{-1}$. Using the 0.3-8 keV luminosity in 2001 and the model for the X-ray spectrum of J1228+441[46] we derive a 2-10 keV luminosity of $\sim 1.7 \times 10^{38}$ erg s$^{-1}$ at 3.8 Mpc, giving log $R_X \sim -2.7$. Since even extremely radio luminous supernova remnants are relatively weak in hard X-rays, it is unlikely that they are the origin of the radio *and* X-ray emission from the nuclear source in Henize 2-10.

iii. Young Supernovae?

Young supernovae having strong interactions with their circumstellar medium and/or a dense ISM can be extremely luminous in both radio and hard X-rays, and can even exceed the luminosities of the nuclear source in Henize 2-10 (e.g. SN 1979C[47,48], SN 1993J[49,50]). Moreover, at some times during their early evolution, their $R_X$ values can also be similar to that of the nuclear source in Henize 2-10. At these early times, radio supenovae are extremely compact with typical sizes of $\lesssim 0.1$ pc[51,52]. Since a previous VLBI study of Henize 2-10[23] detected no sources above 0.1 mJy beam$^{-1}$ (the $5\sigma$ detection limit) with a beam size corresponding to $\sim$0.5 pc $\times$ 0.1 pc, young supernovae with flux densities above 0.1 mJy ($\sim$ Cas A at the distance of Henize 2-10) can be ruled out as the origin of the nuclear source in Henize 2-10.

In principle, more than 15 radio supernovae with individual fluxes beneath the sensitivity limit of the VLBI observations could go undetected yet still add up to the total VLA flux of the nuclear source in Henize 2-10. However, it is extremely unlikely that more than 15 supernovae exist within our VLA beam of $\sim$24 pc $\times$ 9 pc. At these spatial scales, the most likely host for a large number of supernovae would be one or more young massive star clusters, which have typical sizes of $\sim$5 pc. For a generous supernova lifetime of 300 years (i.e. the approximate age of the supernova *remnant* Cas A), a supernova rate of 0.025 per year would be required to have a > 1% chance of there being 15 or more supernovae. This supernova rate exceeds the maximum predicted supernovae rate of a massive super star cluster of $10^6$ M$_\odot$ by more than a factor of six[53]. Such a massive cluster would be optically bright; however, no clusters are observed at the location of the nuclear radio source. While it is possible that the supposed star cluster is obscured by dust, the detection of an H$\alpha$ counterpart to the radio source suggests this is unlikely.

iv. Different origins for the radio and X-ray emission?

We note that the position of the hard X-ray source is also consistent (within $3\sigma$) with nearby young super star clusters (clusters 3 & 4 from Chandar et al. 2003[54]), which are potential hosts of high mass X-ray binaries or recent supernovae. However, the relative astrometry between the VLA and *HST* images allows us to definitively rule out the association of the central radio source with these super star clusters. We therefore consider possible scenarios in which the radio and X-ray emission have different origins.

High mass X-ray binaries are a potential source of the observed hard X-ray emission, but not the radio emission. The expected number of high-mass X-ray binaries in a collection of young super star clusters can be estimated based on observations of the Antennae galaxies[55,56]. Given a total mass of $4 \times 10^5$ M$_\odot$ between the two super star clusters consistent with the position of the hard X-ray source in Henize 2-10, we expect $\sim 6.8 \times 10^{-3}$ X-ray binaries with $L_X \gtrsim 10^{37}\,\mathrm{erg\,s^{-1}}$. At the luminosity of the hard X-ray source in Henize 2-10 ($\sim 2.7 \times 10^{39}\,\mathrm{erg\,s^{-1}}$), the expected number of X-ray binaries drops to $\sim 4 \times 10^{-4}$.

A young supernova that is exceptionally weak in the radio (to be under the VLBI detection limit, see above) is another candidate for producing the hard X-ray emission without producing the radio emission. However, a comparison with the light curves of SN 1993J[50,51] indicates that a supernova with these properties (i.e. $\log R_X < -4.7$) would be extremely young ($<< 1$ year), at an age before the radio luminosity has reached its peak. Given this short timescale, catching such a supernova is unlikely.

The radio emission from the nuclear source in Henize 2-10 could, in principle, be due to multiple supernova remnants. For example, it is possible that more than 15 supernova remnants like Cas A or more than 2 ultraluminous supernova remnants like J1228+441 (see above) could produce the equivalent radio luminosity without producing the observed hard X-ray emission. However, we find this scenario hard to reconcile with the absence of any massive star clusters within the VLA beam of $\sim 24$ pc $\times$ 9 pc, since these objects are the most likely hosts for multiple supernova remnants within such a small area. As mentioned above, the detection of an H$\alpha$ counterpart to the nuclear VLA source suggests extinction is unlikely to account for the lack of a visible star cluster. To illustrate this point further, we note the presence of another compact non-thermal source in Henize 2-10 that is clearly associated with the most luminous super star cluster in the galaxy (i.e. the brightest cluster in Figure 2 and cluster 1 in Chandar et al. 2003[54]). The radio emission from this other source is almost certainly due to one or more supernova remnants. In sum, while it may be possible to contrive a scenario that does not require the presence of an active nucleus in Henize 2-10, the probability of such a scenario is exceedingly low. A low-luminosity AGN is the most natural explanation for the nuclear source in this dwarf galaxy.

**Eddington Ratio of the Low-Luminosity AGN**

Here we consider the accretion power of the low-luminosity AGN in Henize 2-10. For an X-ray luminosity of $2.7 \times 10^{39}$ erg s$^{-1}$ in the 2-10 keV band, the Eddington Ratio is given by $\sim 10^{-4} \times (\kappa/10) \times (2 \times 10^6 \, M_\odot/M_{\rm BH})$, where $\kappa = L_{\rm bol}/L_{\rm X}$ is the 2-10 keV X-ray bolometric correction[57] and $M_{\rm BH}$ is the mass of the black hole in solar masses. Using our best estimate of the black hole mass, $M_{\rm BH} = 2 \times 10^6 \, M_\odot$, and $\kappa = 10$, the Eddington ratio for the black hole in Henize 2-10 is $\sim 10^{-4}$. This value is uncertain by a factor of $\sim 10$ given the error of $\sim 1$ dex in mass. The black hole in Henize 2-10 is currently radiating significantly below its Eddington limit.

Supplementary Table 1. Observations of Henize 2-10

| Telescope | Description |
|---|---|
| VLA | C-band (6.2 cm, 4.9 GHz) radio continuum |
| VLA | X-band (3.5 cm, 8.5 GHz) radio continuum |
| HST | NICMOS F205W, $\sim K$-band (2.1 $\mu$m) |
| HST | NICMOS F190N, Pa$\alpha$ continuum (1.9 $\mu$m) |
| HST | NICMOS F187N, Pa$\alpha$ line (1.87 $\mu$m) |
| HST | ACS F814W, $\sim I$-band (0.8 $\mu$m) |
| HST | WFPC2 F658N, H$\alpha$ line (0.6 $\mu$m) |
| HST | ACS F330W, $\sim U$-band (0.3 $\mu$m) |
| Chandra | ACIS, 0.5-10 keV |

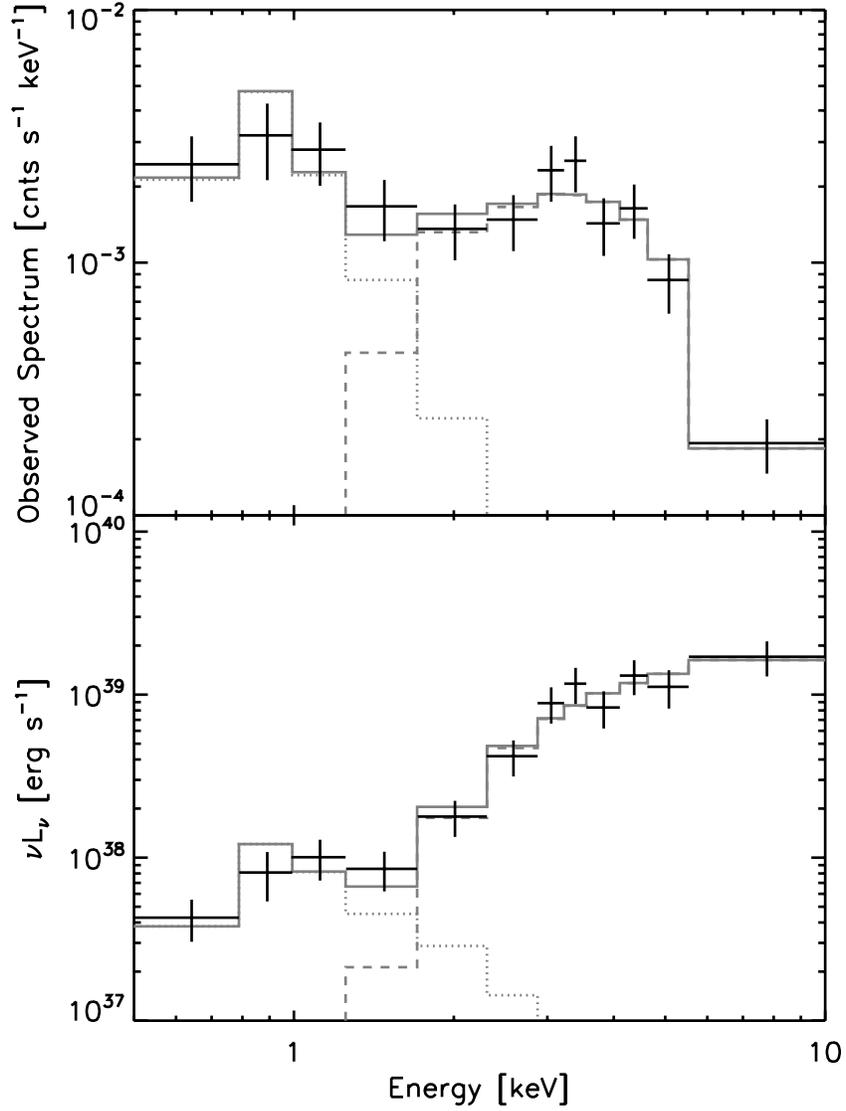

Supplementary Figure 1 - CXO spectrum of the nuclear source in Henize 2-10. The top panel displays the observed spectrum with $1\sigma$ standard deviation errors, normalized to the exposure of the observation. Despite lower effective areas at harder energies ($\gtrsim 2\,\mathrm{keV}$), the majority of photons are hard. The best-fit model, after having been folded through the instrumental response, is displayed in grey with dotted, dashed, and solid lines indicating the soft diffuse gas component, absorbed hard power-law component, and their sum, respectively. In the bottom panel we display the spectrum and best-fit model as observed $\nu L_\nu$, assuming the distance to Henize 2-10. Note that the conversion to $\nu L_\nu$ is model dependent and the effects of neither intrinsic nor Galactic absorption have been removed. The hard power-law component dominates the total power of the source.